\documentclass[12pt]{article}
\usepackage{graphics,psfrag}
\usepackage{amsthm,amssymb,epsfig,amsmath,euscript,array,cite}

\newcounter{multieqs}

\newcommand{\be}{\begin{equation}}
\newcommand{\ee}{\end{equation}}
\newcommand{\eq}[1]{(\ref{#1})}

\newcommand{\bm}[1]{\mbox{\boldmath $#1$}}
\newcommand{\rf}[1]{(\ref{#1})}

\def\bd{\begin{document}}
\def\ed{\end{document}}
\def\nn{\nonumber}
\def\bea{\begin{eqnarray}}
\def\eea{\end{eqnarray}}
\let\bm=\bibitem
\let\la=\label

\def\npb#1#2#3{Nucl. Phys. {\bf{B#1}} #3 (#2)}
\def\plb#1#2#3{Phys. Lett. {\bf{#1B}} #3 (#2)}
\def\prl#1#2#3{Phys. Rev. Lett. {\bf{#1}} #3 (#2)}
\def\prd#1#2#3{Phys. Rev. {D \bf{#1}} #3 (#2)}
\def\cmp#1#2#3{Comm. Math. Phys. {\bf{#1}} #3 (#2)}
\def\cqg#1#2#3{Class. Quantum Grav. {\bf{#1}} #3 (#2)}
\def\nppsa#1#2#3{Nucl. Phys. B (Proc. Suppl.) {\bf{#1A}}#3 (#2)}
\def\ap#1#2#3{Ann. of Phys. {\bf{#1}} #3 (#2)}
\def\ijmp#1#2#3{Int. J. Mod. Phys. {\bf{A#1}} #3 (#2)}
\def\rmp#1#2#3{Rev. Mod. Phys. {\bf{#1}} #3 (#2)}
\def\mpla#1#2#3{Mod. Phys. Lett. {\bf A#1} #3 (#2)}
\def\jhep#1#2#3{J. High Energy Phys. {\bf #1} #3 (#2)}
\def\atmp#1#2#3{Adv. Theor. Math. Phys. {\bf #1} #3 (#2)}

\newcommand{\EQ}[1]{\begin{equation} #1 \end{equation}}
\newcommand{\AL}[1]{\begin{subequations}\begin{align} #1 \end{align}
\end{subequations}}
\newcommand{\SP}[1]{\begin{equation}\begin{split} #1 \end{split}\end{equation}}
\newcommand{\ALAT}[2]{\begin{subequations}\begin{alignat}{#1} #2 
\end{alignat}\end{subequations}}
\def\beqa{\begin{eqnarray}} 
\def\eeqa{\end{eqnarray}} 
\def\beq{\begin{equation}} 
\def\eeq{\end{equation}} 

\def\N{{\cal N}}
\def\sst{\scriptscriptstyle}
\def\thetabar{\bar\theta}
\def\Tr{{\rm Tr}}
\def\one{\mbox{1 \kern-.59em {\rm l}}}

\def\a{\alpha}      \def\da{{\dot\alpha}}  
\def\b{\beta}       \def\db{{\dot\beta}}  
\def\c{\gamma}  \def\C{\Gamma}  \def\dc{{\dot\gamma}}  
\def\d{\delta}  \def\D{\Delta}  \def\ddt{\dot\delta}  
\def\e{\epsilon}        \def\vare{\varepsilon}  
\def\f{\phi}    \def\F{\Phi}    \def\vvf{\f}  
\def\h{\eta}  
\def\k{\kappa}  
\def\l{\lambda} \def\L{\Lambda}  
\def\m{\mu} \def\n{\nu}  
\def\o{\omega}  
\def\p{\pi} \def\P{\Pi}  
\def\r{\rho}  
\def\s{\sigma}  \def\S{\Sigma}  
\def\t{\tau}  
\def\th{\theta} \def\Th{\Theta} \def\vth{\vartheta}  
\def\X{\Xeta}  
\def\z{\zeta}  

\def\na{\nabla}  
\def\cA{{\cal A}} \def\cB{{\cal B}} \def\cC{{\cal C}}  
\def\cD{{\cal D}} \def\cE{{\cal E}} \def\cF{{\cal F}}  
\def\cG{{\cal G}} \def\cH{{\cal H}} \def\cI{{\cal I}}  
\def\cJ{{\cal J}} \def\cK{{\cal K}} \def\cL{{\cal L}}  
\def\cM{{\cal M}} \def\cN{{\cal N}} \def\cO{{\cal O}}  
\def\cP{{\cal P}} \def\cQ{{\cal Q}} \def\cR{{\cal R}}  
\def\cS{{\cal S}} \def\cT{{\cal T}} \def\cU{{\cal U}}  
\def\cV{{\cal V}} \def\cW{{\cal W}} \def\cX{{\cal X}}  
\def\cY{{\cal Y}} \def\cZ{{\cal Z}}

\def\ua{\underline{\alpha}}  
\def\ub{\underline{\phantom{\alpha}}\!\!\!\beta}  
\def\uc{\underline{\phantom{\alpha}}\!\!\!\gamma}  
\def\um{\underline{\mu}}  
\def\ud{\underline\delta}  
\def\ue{\underline\epsilon}  
\def\una{\underline a}\def\unA{\underline A}  
\def\unb{\underline b}\def\unB{\underline B}  
\def\unc{\underline c}\def\unC{\underline C}  
\def\und{\underline d}\def\unD{\underline D}  
\def\une{\underline e}\def\unE{\underline E}  
\def\unf{\underline{\phantom{e}}\!\!\!\! f}\def\unF{\underline F}  
\def\unm{\underline m}\def\unM{\underline M}  
\def\unn{\underline n}\def\unN{\underline N}  
\def\unp{\underline{\phantom{a}}\!\!\! p}\def\unP{\underline P}  
\def\unq{\underline{\phantom{a}}\!\!\! q}  
\def\unQ{\underline{\phantom{A}}\!\!\!\! Q}  
\def\unH{\underline{H}}  
  
\def\As {{A \hspace{-6.4pt} \slash}\;}  
\def\bs {{b \hspace{-6.4pt} \slash}\;}  
\def\Ds {{D \hspace{-6.4pt} \slash}\;}  
\def\ds {{\del \hspace{-6.4pt} \slash}\;}  
\def\ss {{\s \hspace{-6.4pt} \slash}\;}  
\def\ks {{ k \hspace{-6.4pt} \slash}\;}  
\def\ps {{p \hspace{-6.4pt} \slash}\;}   
\def\xs {{x \hspace{-6.4pt} \slash}\;}  
\def\pas {{{p_1} \hspace{-6.4pt} \slash}\;}  
\def\pbs {{{p_2} \hspace{-6.4pt} \slash}\;}  
  
\def\Dh{\hat{D}}
\def\Gh{\hat{G}}
\def\Fh{\hat{F}}
\def\Ph{\hat{P}}
\def\Vh{\hat{V}}  
\def\Xh{\hat{X}} 
 
\def\ah{\hat{a}}
\def\gh{\hat{g}} 
\def\hh{\hat{h}}
\def\xh{\hat{x}}  
\def\yh{\hat{y}}  
\def\ph{\hat{p}}  
\def\xih{\hat{\xi}}  
  
\def\psit{\tilde{\psi}}  
\def\Psit{\tilde{\Psi}}   
\def\Psibt{\tilde{\bar{Psi}}}  

\def\Phit{\tilde{\Phi}}   
\def\Phitb{\overline{\tilde{Phi}}}  
\def\tht{\tilde{\th}}  
\def\lt{\tilde{\l}}
\def\chit{\tilde{\chi}}   

\def\At{\tilde{A}}
\def\Dt{\tilde{D}}
\def\Ft{\tilde{F}}
\def\Qt{\tilde{Q}}  
\def\Rt{\tilde{R}}  
\def\Mt{\tilde{M }}  
\def\Nt{\tilde{N}}   
\def\Xt{\tilde{X}}
 
\def\at{\tilde{a}}  
\def\htt{\tilde{h}} 
\def\st{\tilde{s}}  
\def\ft{\tilde{f}}
\def\gt{\tilde{g}}
\def\pt{\tilde{p}}  
\def\qt{\tilde{q}}  
\def\vt{\tilde{v}}  
\def\nt{\tilde{n}}  
\def\ut{\tilde{u}} 
\def\xt{\tilde{x}} 
\def\yt{\tilde{y}} 
\def\Psit{\tilde{\Psi}}
\def\vphit{\tilde{\varphi}}  

\def\delb{\overline{\partial}}  
\def\thb{\overline{\theta}}
\def\mub{{\overline \mu}}
\def\lamb{{\overline \l}}
\def\psib{{\overline \psi}}
\def\sb{{\overline \sigma}}
\def\xib{{\overline \xi}}
\def\chib{{\overline \chi}}

\def\Phib{\overline{\Phi}}
\def\Lamb{\overline{\Lambda}}

\def\Ab{{\overline A}} \def\Bb{{\overline B}} \def\Cb{{\overline C}}  
\def\Db{{\overline D}} \def\Eb{{\overline E}} \def\Fb{{\overline F}}  
\def\Gb{{\overline G}} \def\Hb{{\overline H}} \def\Ib{{\overline I}}  
\def\Jb{{\overline J}} \def\Kb{{\overline K}} \def\Lb{{\overline L}}  
\def\Mb{{\overline M}} \def\Nb{{\overline N}} \def\Ob{{\overline O}}  
\def\Pb{{\overline P}} \def\Qb{{\overline Q}} \def\Rb{{\overline R}}  
\def\Sb{{\overline S}} \def\Tb{{\overline T}} \def\Ub{{\overline U}}  
\def\Vb{{\overline V}} \def\Wb{{\overline W}} \def\Xb{{\overline X}}  
\def\Yb{{\overline Y}} \def\Zb{{\overline Z}}  

\def\fb{{\overline f}}
\def\gb{{\overline g}}
\def\mb{{\overline m}}
\def\lb{{\overline l}}
\def\yb{{\overline y}}

\def\ba{{\bf a}} 
\def\bk{{\bf k}}  
\def\bl{{\bf l}}  
\def\bp{{\bf p}}  
\def\bq{{\bf q}}  
\def\br{{\bf r}}
\def\bt{{\bf t}}
\def\bu{{\bf u}}
\def\bv{{\bf v}}
\def\bx{{\bf x}}  
\def\by{{\bf y}}  
\def\bR{{\bf R}}  
\def\bV{{\bf V}}

\def\bone{{\bf 1}}  

\def\va{{\vec a}}
\def\vk{{\vec k}}
\def\vp{{\vec p}}
\def\vq{{\vec q}}
\def\vx{{\vec x}}
\def\vy{{\vec y}}
\def\vu{{\vec u}}
\def\vv{{\vec v}}

\def\vs{{\vec \sigma}}
\def\vtau{{\vec \tau}}

\newcommand{\ov}[1]{\overrightarrow{#1}}
  
\def\d{\delta}\def\D{\Delta}\def\ddt{\dot\delta}  
  
\def\pa{\partial} \def\del{\partial}  
\def\xx{\times}  
\def\uno{\mbox{1 \kern-.59em {\rm l}}}    
  
\def\trp{^{\top}}  
\def\inv{^{-1}}  
\def\dag{{^{\dagger}}}  
\def\pr{^{\prime}}  
  
\def\rar{\rightarrow}  
\def\lar{\leftarrow}  
\def\lrar{\leftrightarrow}  
  
\newcommand{\0}{\,\!}      %this is just NOTHING!  
\def\one{1\!\!1\,\,}  
\def\im{\imath}  
\def\jm{\jmath}  
  
\newcommand{\tr}{\mbox{tr}}  
\newcommand{\slsh}[1]{/ \!\!\!\! #1}  
  
\def\vac{|0\rangle}  
\def\lvac{\langle 0|}  
  
\def\hlf{\frac{1}{2}}  
\def\ove#1{\frac{1}{#1}}  

\def\Box{\square}  
\def\ZZ{\mathbb{Z}}  
\def\bb#1{{\bf #1}}  
\def\bcomment#1{}  
\def\bfhat#1{{\bf \hat{#1}}}  
\def\VEV#1{\left\langle #1\right\rangle}  

\newcommand{\ex}[1]{{\rm e}^{#1}} \def\ii{{\rm i}}  

\newcommand{\lrbrk}[1]{\left(#1\right)}
\newcommand{\sfrac}[2]{{\textstyle\frac{#1}{#2}}}
 
\def\stw{{\sqrt{2}}}

\def\rf {{\rm f}}
\def\ri {{\rm i}}
\def\rs {{\scriptscriptstyle \rm S}}
\def\rt {{\scriptscriptstyle \rm T}}

\def\rQ {{\scriptscriptstyle \rm \cQ}}
\def\rR {{\scriptscriptstyle \rm \cR}}

\def\cQb{{\cal \Qb}}
\def\cRb{{\cal \Rb}}
\def\cWb{{\cal \Wb}}

\def\fd {{\rm N}}
\def\afd {{\overline{\rm N}}}

\def \II {I\hspace{-.1em}I\hspace{.1em}}
\def \IIA {\mbox{\II A\hspace{.2em}}}
\def \IIB {\mbox{\II B\hspace{.2em}}}
\def \gs {g^s}
\def \ls {\lambda^s}

\begin{document}

\hfill{hep-th/0602054}

\vspace{20pt}

\begin{center}

{\Large \bf  Time-dependent AdS/CFT Duality and Null Singularity
}
\vspace{30pt}

{\bf Chong-Sun Chu$^1$, Pei-Ming Ho$^2$}

\vspace{15pt}
{\small \em
\begin{itemize}
\item[$^1$]
Centre for Particle Theory and Department of Mathematics,
University of Durham, Durham, DH1 3LE, UK.
\item[$^2$]
Department of Physics, National Taiwan University, Taipei, Taiwan, R.O.C.
\end{itemize}
}

\vskip .1in {\small \sffamily chong-sun.chu@durham.ac.uk, 
pmho@phys.ntu.edu.tw}

\vspace{50pt}
{\bf Abstract}
\end{center} 
 
We consider AdS/CFT correspondence for time-dependent \II B backgrounds in
this paper. The supergravity solutions we construct are supersymmetric
pp-waves on AdS and may have null singularity in the bulk. The dual gauge
theory is also constructed explicitly and is given by a time-dependent
supersymmetric Yang-Mills theory living on the boundary. Apart from the
usual terms that are dictated by the geometry, our gauge theory action
features also a time-dependent axion coupling and a time-dependent gauge
coupling. Both of which are necessary due to the presence of a nontrivial
dilaton and axion profile in the supergravity solution. The proposal is
supported by a precise matching in the symmetries and functional
dependence on the null coordinate of the two theories. As applications, we
show how the bulk Einstein equation may be reproduced from the gauge
theory. We also study and compare the behaviour of the field theory
two-point functions. We find that the two-point function computed by using
duality is different from that by doing a direct field theory computation.
In particular the spacetime singularity is not seen in our gauge theory
result, suggesting that the spacetime singularity may be resolved in the
gauge theory.

\setcounter{page}0
\newpage

\section{Introduction}

The understanding of the fundamental nature and quantum properties of
spacetime is one of the most important questions for theoretical high
energy physics. String theory is a consistent theory of quantum gravity. As
such, string theory should be able to provide a satisfactory resolution for
outstanding problems such as the entropy of a black hole, spacelike
singularity inside the horizon of a black hole and the big bang singularity
in the early universe. To achieve this, it is necessary to understand
string theory on time-dependent backgrounds \cite{hor1}. Unfortunately time
dependent backgrounds are difficult to work with in general. An exceptional
class of models which is simple enough to work with and yet may provide a
useful approach is the time-dependent orbifold model discussed first by
Horowitz and Steif \cite{hor2}. In particular, the singularity properties
of the string S-matrix and the stability of the background against
gravitational backreaction has been examined
%p7
\cite{LMS1,Lawrence,FM,HP,Berkooz,Cornalba} and it is shown that a proper
understanding and treatment of the gravitational backreaction is crucial to
establish reliable string results. An understanding of string theory in
time-dependent background is still beyond the scope of traditional
perturbative string theory.

Recently, powerful nonperturbative formulations of string theory,
namely matrix theory \cite{mat,mat-review}
and AdS/CFT correspondence \cite{mal,mal-review},
have been proposed and put forward. It is natural to apply
these ideas in the studies of time-dependent backgrounds, and try to say
something about the nature and fate of the spacetime singularity.
Holographic description of time-dependent backgrounds via the
AdS/CFT correspondence was considered in, for example,
%c7
\cite{Brecher,adscft1,adscft2,adscft3} and 
more recently in \cite{adscft4} where
a big crunch cosmology was constructed and the dual field theory
description was examined. In an interesting proposal \cite{csv},
a simple supersymmetric time-dependent background
with a null singularity is the holographic dual of a matrix string theory.
The time-dependence of the background is proposed to be encoded through
the time-dependent coupling of the gauge theory. Extensions and related
works can be found in \cite{matrix1}. Holographic description of
cosmology in terms of matrix theory can be found in \cite{matrix2}.

As is usual in AdS/CFT correspondence, supersymmetry can be expected to play
an important role. The existence of supersymmetry allows us a better
control over the string/supergravity background and over the quantum
and nonperturbative behaviour of the field theory. In the construction
of \cite{adscft4}, the supergravity solution was constructed for a class
of boundary conditions preserving the asymptotically AdS symmetries.
However supersymmetry is not preserved in
that construction. On the
other hand, in the proposal \cite{csv}, although the background is half BPS,
the supersymmetry of the dual matrix string theory is 
%cc2 not explicit.
broken due to the presence of finite lightcone momentum.

Supersymmetric time-dependent background has been considered
in earlier works. In particular in \cite{Brecher}, a time-dependent
deformation of the pp-wave geometry has been constructed and the AdS/CFT
correspondence considered. Although the background in these works has
pp-singularities \cite{podolsky}, these singularities are situated at
the horizon ($u=\infty$) 
and as a result one can  expect that their effects 
will be red-shifted away and will not
show up in the dual field theory \cite{Chamblin}. This is indeed the case
as confirmed by the calculations of \cite{Brecher}. Moreover these
singularities are spacelike and are of different kind from the
singularities one would like to study in cosmology.

In this paper, we follow the line of AdS/CFT correspondence. We
construct supersymmetric \IIB backgrounds which carry
nontrivial dilaton and axion
profiles.
The dual gauge theory is also 
constructed explicitly. The dual theory features a time-dependent gauge coupling 
and a time-dependent axion coupling
and is $\cN=2$ supersymmetric,
preserving the same amount (1/4)
of supersymetries as the supergravity (SUGRA) solution. 
Our goal is to use
the dual gauge theory to try to get a better understanding  about string
theory in non-static spacetime and about the properties of spacetime
singularity.  On this, we notice  that by allowing
nontrivial scalars profile,
our SUGRA solution can admit not just the usual
pp-singularities (which are irrelevant for our studies), but also
cosmological type (to be precise, null-like)
singularities.
Moreover since these
singularities are
situated at a constant $x^+$,
their presence can in principle be detected by quantities computed
in the dual field theory. For this purpose, we  compare 
the field theory two-point functions computed from two different
methods: one computed using  the bulk-boundary propagator
and the other computed directly from the field
theory. We find that the two don't agree. In particular the SUGRA
result is sensitive to the singularity of the spacetime, while the
gauge theory result does not see the singularity. That the results 
differ is not surprising since the SUGRA result 
is valid in the regime where the t'Hooft
coupling is  large, while the field theory result 
is valid when the t'Hooft coupling is
small (in our case it is zero since our computation is performed at
the free level). As argued by \cite{shenker}, a field
theory computation at weak coupling should still be valuable in
capturing the singularity behaviour of the spacetime, even if only
qualitative. Therefore our field theory 
result suggests that the spacetime singularity as seen at
the SUGRA level could be resolved by $\a'$ effects of string theory.
A similar suggestion has also been reached  at \cite{shenker}.

The paper is organized as follows. In section 2, we present our
construction of the time-dependent solution in type \IIB supergravity. In
section 3 we present our construction of the time-dependent supersymmetric
gauge theory and give justifications.
In section 4, we apply the duality and try to use gauge theory to learn
about the dual spacetime properties. In particular we compute the 
two-point functions and study their behaviour in relation to spacetime 
singularity structure. 
We also show how the finiteness of the vacuum expectation value (VEV)
of the energy momentum
tensor may allow us to derive the Einstein equations in the bulk.
We end in section 5 with a few remarks and further discussions.

\section{Time-dependent Deformation of AdS Background}

We are interested in generalizing the original
Maldacena AdS/CFT correspondence to one where
the asymptotic AdS space is deformed to
have nontrivial time-dependence.
This will be a new class of duality different from
existing studies of time-dependent processes
in AdS/CFT correspondence.
We will work with the extension of the AdS/CFT
correspondence for the type \II B case. It should be straightforward to
generalize our analysis to the 11 dimensional $AdS_4 \times S^7$ or
$AdS_7 \times S^4$ cases.

In this paper, we will consider solutions where the NSNS and RR 2-form
potentials vanish. Generalization to include nontrival 2-form potentials
is possible. 
In Einstein frame, the equations of motion for the bosonic fields are
\cite{jhs}
\bea
&\na_M(e^{2\phi} \na^M \chi) = 0, \label{eom1}\\
&\na^M \na_M \phi + e^{2\phi} \del^M \chi \del_M \chi = 0,
\label{eom2}\\
& e F_{M_1 \cdots M_5} =  
\varepsilon_{M_1 \cdots M_5}{}^{ L_1 \cdots L_5} F_{L_1 \cdots
  L_5},\label{eom3} \\
& R_{MN} = \frac{1}{2} \del_M \phi \del_N \phi + \frac{1}{2} e^{2\phi}
\del_M \chi \del_N \chi + \frac{1}{6} F_{L_1 \cdots L_4 M} 
F^{L_1 \cdots L_4}{}_N . \label{eom4} 
\eea
Here $M,N = 0,1,\cdots, 9$. $e^2$ is the determinant of the metric
$g_{MN}$. 

\subsection{The solution}
 
To solve \eq{eom1} -\eq{eom4} with non-trivial time-dependence, we will
need to impose an appropriate ansatz. Since a time-dependence
in the $S^5$ part of the metric is harder to interpret in the dual
gauge theory, we will restrict ourselves in this paper to
deformations only in the $AdS_5$ part of the metric.
We start with the following ansatz ($i =2,3$),
\bea
ds^2 &=& \frac{R^2}{u^2} \left(
- k^2(x^+)dx^+ dx^-  + h(u, x^-, x^+, x^i) (dx^+)^2
+ (d x^i)^2 + du^2 \right)
+ R^2 d\Omega_5^2, \label{ansatz-metric} \\
F_{\m\n\r\l\s} &=& \frac{1}{R} \varepsilon_{\m\n\r\l\s},
\quad
F_{abcde} = \frac{1}{R} \varepsilon_{abcde},
\quad \mbox{for the $AdS_5$-like and $S^5$ part respectively},
\label{ansatz-F}
\eea
\be
\phi = \phi(x^+), \quad \chi =\chi(x^+), \label{ansatz-scalar}
\ee
where the tangent space components of the 5-form is
given in  \eq{ansatz-F} 
and the functions $k, h, \phi, \chi$ are defined over the whole real axis.
A couple of remarks follow.
1. We have chosen not to include any $x^+$
dependence in  front of the $du^2$ term as we would like to maintain the
interpretation of $u$ as the holographic energy scale, and a time
dependence in it would render it difficult for such an interpretation.
2. By a redefinition of $x^+$,
one can normalize $k$ to 1 generically. Here we leave
the possibility of an explicit $x^+$ dependence in $k$.
This will allow to include singular solutions which are
geodesically incomplete (see section \ref{incomplete})
without having to introduce singularities in the metric.
3. Similar solutions which are asymptotically
AdS were studied before
%p7
\cite{Chamblin,Brecher,Kumar,Sfetsos}.
The major feature which distinguishes our solution
from existing solutions is that nontrivial
(time-dependent) dilaton and axion configurations
are turned on in our solution
%p7
\footnote{In \cite{Sfetsos},
the dilaton is turned on for
other branes but not the D3-brane
which is the one related to our solutions}
%p8
\footnote{
We thank Arkady Tseytlin for pointing out
to us that our solution can be related to those
in \cite{Tseytlin} by a chain of duality
relations and with dilaton
and axion fields turned on.}.
This allows us to include solutions  with spacetime singularity in the bulk.

We now consider the equations of motion. 
The self-duality condition \eq{eom3} is satisfied by our ansatz. 
Also the equations \eq{eom1} and \eq{eom2} are trivially satisfied 
since the scalars depend only on $x^+$.
For the Einstein equation \eq{eom4}, we note that
the metric \eq{ansatz-metric} has a modified Ricci tensor whose
nonvanishing components are given by 
\bea
&& R_{uu} =R_{22} = R_{33} =  - \frac{4}{u^2}, \\
&&R_{+-} =   \frac{2 k^2}{u^2} - \frac{1}{k^2}\del_-^2 h, 
\qquad R_{+I} = - \frac{1}{k^2} \del_- \del_I h, \quad I =2,3,u,\\ 
&&R_{++} = - \frac{4h}{u^2} + \frac{3\del_u h}{2u} 
+\frac{2}{k^4}  h \del_-^2 h
- \frac{1}{2} ( \del_u^2 h + \del_2^2 h + \del_{3}^2 h ).
\eea
It is easy to see that the Einstein equation \eq{eom4}
can be satisfied if $h$ is of the form
\be \label{h-gen}
h  = h_0 + h_- x^- +\sum_i \left( k_i x^i+ h_{ij} \, x^i x^j \right)
+ l_u u^2 + (l_0 + \sum_i l_i \, x^i)u^4,
\ee
where all the coefficients $h_0 ,\cdots, l_3$ above 
are functions of $x^+$; and they satisfy
\be
\frac{1}{2} (\phi')^2 + \frac{1}{2} e^{2\phi}(\chi')^2 =
 - h_{22} -h_{33}+ 2l_u,
\ee
which follows from the (++)-component of the Einstein equation.
In \cite{Brecher}, a nondilatonic background was considered and 
solution with nonzero  $h_{22} =h_{33} =l_u/2$ and $l_0$ was constructed.
As is usual in the AdS/CFT correspondence
\footnote{We recall that in the AdS/CFT correspondence, solutions in the bulk of
AdS are matched with either physical states or background deformations of 
the boundary theory according to their radial dependence.
%Supergravity solutions with different boundary values correspond to
%different gauge theories.
},
adding $u$-dependent parts in $h$ that do not change the value of $h$ 
on the boundary $u=0$ corresponds to the same boundary theory 
in an excited state.
%Therefore these deformations
%p5 corresponds
%correspond to an excited state in
%the dual field theory. In fact it
It has been argued in \cite{Brecher} that
this background is dual to a field theory
with a constant lightcone momentum density.
In this paper we consider another type of
deformation which changes the boundary metric nontrivially. 
For this purpose,
it is sufficient to ignore the $u$-dependence in \eq{h-gen} 
and consider the ansatz
\be \label{h-gen1}
h = h_0 + h_- x^-  
+\sum_i \left( k_i x^i+ h_{ij} x^i x^j \right).
\ee
It is now straightforward to show that
by performing a coordinate transformation of the form
\bea
&& dx^+ = f(\xt^+) d\xt^+, 
\quad x^i = \xt^i + a^i(\xt^+),
\quad u = \ut,\\
&& \xt^- =f(\xt^+) \left(
x^- + b_i(\xt^+) \xt^i  \right) +
\tilde{\a}(\xt^+),
\eea
one can turn $h_0, h_-, k_i$ to zero. Therefore it is 
sufficient to consider 
\be \label{h-sugra}
h = h_{ij}(x^+) x^i x^j 
\ee 
for the metric and the Einstein equation reads 
\be \label{cond-red}
\frac{1}{2} (\phi')^2 + \frac{1}{2} e^{2\phi}(\chi')^2 =
 - h_{22} -h_{33}.
\ee 
Since $h$ is independent of $x^-$, our metric admits a null Killing
vector $\xi = \del_-$ 
and corresponds to a pp-wave in AdS. 
Time-dependent solutions with a null Killing
vector has also been constructed in string theory, see for example
\cite{null} for some recent discussions.

The metric \eq{ansatz-metric} is written in the 
``Brinkman form", one can bring it to
the``Rosen form" 
\be \label{metric-diag}
 ds^2 = \frac{R^2}{u^2}\left( -k^2(y^+) dy^+ dy^-  +   
\sum_{ij} a_{ij}(y^+) dy^i dy^j  + du^2\right) + R^2
d\Omega_5^2
\ee
by performing a coordinate transformation.
In this form, the metric depends only on the coordinate $y^+$.
Unlike $x^+$ above, $y^+$ is null.  The
coordinate transformation is  complicated in general. 
However it simplifies in the particular case of $h_{23}=0$,
which is equivalent to $a_{23}=0$.  Writing $a_{ij} = M^2_i \d_{ij}$, 
the two coordinate systems are then related by the  transformation
\be
x^+ =y^+, \quad x^i = M_i(x^+) y^i, \quad x^- =y^-+
\sum_{i}\frac{1}{k^2 }M_i'M_i(y^i)^2
\ee
and the functions $h_i$ are related to $ M_i$ by
\be \label{h}
h_i= \frac{M''_i }{M_i} -  \frac{2k' M_i'}{k M_i } , \quad i=2,3.
\ee
The Ricci curvature for the modified AdS part  $(M,N =+,-,2,3,u)$ is
given by,
\be
R_{MN} =  - \frac{4}{u^2}g_{MN} + R_{++} \d_{M+} \d_{N+}, 
\ee 
where $R_{++}$ is given by an expression involving $a_{ij}$
and their derivatives up to the second order. The expression is a little
more complicated and we don't record it here. Again the $(++)$-component
of the Einstein equation gives 
$\frac{1}{2} (\phi')^2 + \frac{1}{2} e^{2\phi}(\chi')^2= R_{++}$
similar to \eq{cond-red}, and presents a constraint among 
$a_{ij}, \phi$ and $\chi$. For the particular case of $h_{23}=a_{23}=0$,
we have
\be \label{R-nn}
R_{++} = -\sum_{i=2,3} 
\left( \frac{M_i''}{M_i} - \frac{2k'M_i'}{k M_i} \right).
\ee

We note  that our supergravity solution  
is invariant under a scaling transformation of the following form:
\be \label{scaling-sugra}
u \to \l u, \quad x^+ \to x^+, \quad x^- \to \l^2 x^-, 
\quad x^i \to \l x^i.
\ee
This  scaling symmetry will play a significant role in 
the dual gauge theory. 

Before we move on, 
we remark that the above construction can be obtained from a suitable
near horizon limit of D3-brane solutions.
Consider an ansatz of the following form
\bea
&ds^2 = H^{-1/2} \big(
- k^2(x^+)dx^+ dx^-  + h(r,x^+, x^i) (dx^+)^2
+ (d x^i)^2  \big)
+ H^{1/2}(d r^2 + r^2 d\Omega_5^2), \quad
 \label{genansatz-m} \\
&F_5 = G_5 + *G_5, \quad \mbox{where} \quad
G_5 = \frac{k^2}{4} d(H^{-1})\wedge dx^+\wedge dx^-\wedge dx^2\wedge dx^3,
\label{genansatz-F}
\eea
\be 
 \phi = \phi(x^+), \quad \chi =\chi(x^+), \label{genansatz-s}
\ee
where
\be
H = 1 + \frac{R^4}{r^4}, \quad \mbox{for arbitrary constant $R$}
\ee
One can check that the type \II B SUGRA
equations of motion are satisfied if
\be
-\frac{1}{2}H^{-1} (\del_r^2 h + \frac{5}{r} \del_r h)
-\frac{1}{2} (\del_2^2 +\del_3^2) h =
\frac{1}{2} (\phi')^2 + \frac{1}{2} e^{2\phi}(\chi')^2.
\ee
This can be satisfied if $h$ takes the form \eq{h-sugra} and obeys
\eq{cond-red}.
Our solution above for pp-wave in AdS can be obtained from this
solution by taking a near horizon limit $r\to 0$ and identify
$u=R^2/r$. The solution \eq{genansatz-m}-\eq{genansatz-s} describes a
stack of  D3-branes with a pp-wave on it.

\subsection{Supersymmetry}

We  now show that our solution preserves 1/4 of 
the \IIB supersymmetry \cite{jhs}.
The  preserved supersymmetry is determined by the
vanishing of the variations of the spinor $\d \l$ and the
gravitino $\d \psi_M$. For our solutions where
the RR and NSNS 2-form potentials are zero, we have
\be\label{susy-cond} 
0= \d \l = i \gamma^M \e^* P_M, \quad  \quad
0= \d \psi_M = (D_M - \frac{i}{2} Q_M) \e +
\frac{i}{480} F_{P1 \cdots P_5}\c^{P_1 \cdots P_5} \c_M  \e .
\ee
Here $P_M$ and $Q_M$ are currents defined by 
$P_M= -\varepsilon_{\a\b}V_+^\a\del_M V_+^\b$, 
$Q_M = -i \varepsilon_{\a\b}V_-^\a\del_M V_+^\b$, where
$V^{\a=1,2}_{\pm}$ is a $2\times 2$ matrix of the scalar fields which
satisfies $\varepsilon V_-^\a V_+^\b =1$ and 
$V_-^\a V_+^\b - V_+^\a V_-^\b = \varepsilon^{\a\b}$. 
In \eq{susy-cond},  $D_M = \del_M  + \frac{1}{4} \o_M{}^{AB} \C_{AB}$
is the covariant derivative, $\o_M{}^{AB}$
is the spin connection matrix. 
$\C_A$ and $\c_M= E_M{}^A \; \C_A $  are 
the $\Gamma$ matrices in the tangent space and coordinate basis respectively:
$\{ \c_M, \c_N \} = 2 g_{MN}, \{ \C_A, \C_B \} = 2 \eta_{AB}.$

To verify the supersymmetry, it is convenient to perform a change of coordinate
$ u = e^{-r}$ 
and use the orthonormal basis
\be
E^r = dr, \qquad E^+ = e^r  dx^+, \qquad E^- = e^r(k^2 dx^- - h  dx^+),
\qquad E^i = e^r dx^i, \quad i=2,3.
\ee
We are choosing the units so that $R=1$ for simplicity. 
The metric takes the form
\be
ds^2 = \eta_{AB} E^A E^B, \quad\mbox{where} \quad 
\eta_{+-} = \eta_{-+} = -1/2,\quad
\eta_{22}=\eta_{33} = \eta_{rr}=1.
\ee
The nonvanishing components of the spin connection $\o^{AB}$ 
($dE^A + \o^A{}_B E^B = 0$) are given by
\bea \label{spin}
&& \o^{+r} = - \o^{r+} = E^+,\quad 
\o^{-r} = - \o^{r-} = E^- -  \del_r h E^+, \quad 
\o^{ir} = - \o^{ri} = E^i,\nn\\
&& \o^{i-} =-\o^{-i} =   e^{-r} \del_i h E^+,\quad
\o^{+-} = - \o^{-+} =  \frac{2 e^{-r}}{k^2}( \del_-h + 2 k k') E^+.
\eea
Here we have given the results for the general case of 
the metric \eq{ansatz-metric} where  $h$ is allowed to  
depend on $r$ and $x^-$, although this is not the case for 
the solutions we construct in this paper.

Now  we examine the supersymmetry conditions 
\eq{susy-cond} for our solution.
Since the scalars  in our solution are functions of $x^+$, 
the only nonvanishing components of $P_M$, $Q_M$
are $P_+, Q_+$. 
The condition $\d \l=0$  reads
\be \label{susy-cond3}
\c^+ \e = \c_- \e =  0, \qquad \mbox{or, equivalently}\qquad
\C^+ \e= \C_- \e =0.
\ee
Due to this condition, the modification of $k$ and 
$h$ in the spin connection \eq{spin} never appears 
since a $\C_-$ is always attached and  will give zero 
upon hitting $\e$. As a result, the equation \eq{susy-cond}
reads the same as in the undeformed $AdS_5 \times S^5$ case.
The $AdS_5$ part of it gives
\bea
&&\del_- \e =0,\qquad \qquad
\left(\del_+ - \frac{i}{2} Q_+ - \frac{e^r}{2}\C_+(1-\C_r) \right) \e =0,\\ 
&&(\del_r -\frac{1}{2} \C_r) \e =0,\quad
\left(\del_i - \frac{e^r}{2}\C_i(1-\C_r) \right) \e =0.
\eea
These are satisfied if
\be \label{poin}
\e = e^{r/2} \e^+_0, \quad \mbox{where}\quad 
\C^r \e^\pm_0 = \pm \e^\pm_0
\ee
for the constant spinors $\e^\pm_0$. 
This is just the usual Poincare supersymmetries of $AdS$, but
with the extra condition
\eq{susy-cond3} imposed.
It is easy to see that the ``AdS-supersymmetry'' 
$\e= (e^{r/2} +e^{-r/2} \xs )\e^-_0$  for the standard $AdS_5\times S^5$ 
is broken since it is not compatible with \eq{susy-cond3}.  
Thus our solution   
preserves 8 supersymmetries of the form $\e= e^{r/2} \e^+_0$ satisfying 
$\Gamma^+ \epsilon = (1-\Gamma^r) \e = 0$.

In the above we have assumed that, given that the metric is nontrivially
dependent on $x^+$, the scalars also depend on $x^+$ nontrivially.
However there is an interesting exception. Consider the case with
\be
h_{22} = -h_{33}.  
\ee
In this case, the dilaton and axion are constant. Although the currents
are now $P_M= Q_M =0$  and there is no need to impose  $\gamma^+
\epsilon =0$ in order for $\d \l =0$, this condition is needed in
order to solve $\d \psi_M=0$. The  solution is 
1/4 BPS as before.

The next simplest example of our solutions is
a linear dilaton background
\be
\phi = \sqrt{-2(h_{22}+h_{33})} x^+, \qquad \chi = 0,
\ee
where $h_{22}+h_{33}$ is a constant.

\subsection{Singularity: geodesic incompleteness} \label{incomplete}

An interesting feature of our supergravity solutions is that 
all gauge invariant quantities constructed out of 
the curvature tensor and the metric are regular.
This does not mean that there is no singularity of any kind in our
solutions. For example, a divergence in the dilaton or axion is still a
singularity. This happens whenever the
component $R_{++}$ of the Ricci tensor has a singularity according to
\eq{cond-red}.
In general a spacetime is singular if it is geodesically incomplete and
cannot be embedded in a larger spacetime \cite{singular}. 
We will now demonstrate that our solutions include situation where
there can be a singularity in the bulk. This is in contrast to those
solutions constructed in e.g. \cite{Brecher,podolsky,Chamblin}.
For the following analysis, it is more convenient to 
consider the Rosen form (\ref{metric-diag}) of the metric. 

We first claim that if we choose $k = 1$ by
redefining the coordinate $y^+$, then 
a singularity in $R_{++}$ also signifies geodesic incompleteness.
The reason is that the curve defined by
\be
y^+ = \l, \qquad y^M = \mbox{constant} \quad \forall \quad M \neq +
\label{geod}
\ee
is a geodesic with the affine parameter $\l \in \bR$
\footnote{The  lines 
$y^- = \l, y^M = \mbox{constant}$ for all $M \neq -$ ,
also form a family of null geodesics. Since our metric is 
independent of $y^-$, it is not very interesting to follow the flow of this 
kind of geodesics.}.
On this geodesic,
\be
R_{MN} \frac{dy^M}{d\l} \frac{dy^N}{d\l} = R_{++}
\ee
is invariant under general coordinate transformations.
Thus a divergence of $R_{++}$ implies that the geodesic
has to be terminated there
and the geodesic is incomplete. A simple example is:
\be \label{eg-1}
\phi = \pm 2\sqrt{\a (1-\a)}\log(y^+),  \quad \chi = 0,
\quad M_2=  M_3 = (y^+)^\a ,
\quad a_{23}=0,\quad k=1.
\ee
Another example is
\be \label{eg-2}
\phi = \phi_0,\quad \chi = \pm 2\sqrt{\a (1-\a)} e^{-\phi_0}\log(y^+),
\quad M_2=  M_3 = (y^+)^\a ,
\quad a_{23}=0,\quad k=1.
\ee
where $y^+>0$ and $0< \a< 1$ in both cases
and $\phi_0$ is a constant. From \eq{R-nn}, 
we have a singularity at $y^+=0$ 
\be
R_{++} = \frac{2 \a (1-\a)}{(y^+)^2}.
\ee
Note that there is a vanishing scale factor resembling the big bang.

Via a coordinate transformation of $y^+$,
which results in a nontrivial function $k$ in the metric,
it is possible to push the singularity to the coordinate infinity,
so that all the fields in our solution appear to be smooth functions.
For example, 
the above singularity at $y^+ = 0$ may be pushed to the infinity
$y^+_{\mbox{\small new}} \rightarrow - \infty$
in terms of the new coordinate $y^+_{\mbox{\small new}} = \log y^+$.
Simultaneously we have $k^2 = 1 \rightarrow k^2 = e^{y^+_{\mbox{\tiny new}}}$.
This is why we claimed earlier that by allowing nontrivial 
$k$ to appear in the $g_{+-}$ component of our metric,
one has the possibility to include singular spacetime 
which is geodesically incomplete
even if all the fields (the dilaton, axion and
all functions defining the metric) are
seemingly regular functions.

In general, 
let us consider  the geodesic equation for the case of a nontrivial $k$
\be
\frac{d^2 y^+}{d \l^2} + \C^+_{++} \frac{dy^+}{d \l}\frac{dy^+}{d \l}
=0,\quad \mbox{where $\C^+_{++}= 2 k'/k$}.
\ee
Up to a constant normalization of the affine parameter $\l$, this gives
\be \label{x-l}
\frac{dy^+}{d\l} = 1/k^2.
\ee
This implies a monotonic relation between $\l$ and $y^+$  and 
could demand $\l$ to be terminated at a finite value.
Moreover  the gauge invariant curvature 
defined along the geodesic is
\be \label{Rll}
R_{\l\l}(\l) \equiv R_{\m\n} \frac{dy^\m}{d\l} \frac{dy^\n}{d\l}
= \frac{1}{k^4} R_{++} \Big|_{y^+ =y^+(\l)} .
\ee
Even if all scalar profiles are smooth and so $R_{++}$ is
nonsingular for all $y^+ \in \bR$,
$R_{\l\l}$ can be singular at finite $\l$
if $k$ approaches to zero somewhere.
In this case the spacetime is geodesic incomplete 
and has a curvature singularity.
On the other hand, if we choose the coordinates with $k = 1$, and all
the fields are smooth functions, the solution is free from any
singularity in the bulk.
(At the horizon $u= \infty$,
there still exists pp-singularity as discussed above.)
%cc1 but it is infinitely far away from the bulk.)

In the above we have considered the metric in the Einstein frame.
The string metric $\gs_{MN}$ differs from the Einstein metric $g_{MN}$
by a factor depending on the dilaton field
\be
g^{(s)}_{MN} = e^{\phi/2} g_{MN}.
\label{metrics}
\ee
Since the dilaton is in general a nontrivial function,
it may happen that a singularity in the Einstein frame
disappears in the string frame.
This is a very interesting situation because
the string is now coupled to a nonsingular metric and if there is no
other singularity, e.g. in the dilaton, then
the theory should be well defined.
Taking advantage of this, a nonperturbative matrix string formulation
has been proposed recently to describe a null cosmological singularity in
the Einstein frame while in the string frame it is a flat metric
with a linear dilaton background \cite{csv}.

Interestingly, our solutions also include geometry of this kind. An
explicit example is that the Einstein frame metric is given by
\eq{metric-diag} with $a_{ij} = \d_{ij} M^2$ and 
\be
k^2 =M^2 = e^{-\phi/2}= \Big(\frac{y^+}{y^+_0}\Big)^{2/3}.
\ee
The affine parameter is $\l =3 (y^+)^{5/3}/5$ and the
gauge invariant curvature is  $R_{\l\l} =
\frac{8}{9}(\frac{y^+}{y^+_0})^{-\frac{10}{3}}$.
The metric is singular at $y^+=0$ and corresponds to
the $\l=0$.
The corresponding string metric is just the undeformed $AdS_5 \times S^5$  and
is regular. Other examples are also possible. 

In conclusion we have shown that our class of solutions is general
enough to include both regular and singular spacetime. Moreover it includes
spacetimes that are singular in the Einstein frame but regular in the
string frame. This kind of
spacetime is of interest for the studies of big bang cosmology. In
the next section, we give a candidate supersymmetric gauge theory that we
propose to be dual to our supergravity solution in general. We emphasis that 
this includes also the singular case.

\section{Time-dependent Supersymmetric Yang-Mills Theory}

Due to the form of our SUGRA metric, it is clear that the boundary manifold 
is equipped with a natural conformal structure but not a natural metric. 
According to \cite{witten-yau},
it is plausible that
there is a correspondence between the
conformal theory on the boundary and quantum gravity in the bulk. However the 
precise form of the boundary theory was not given. 
In this section, we will construct the dual theory directly. 
The dual theory we construct is a time-dependent Yang-Mills theory with
precisely the same amount of functional dependence on $x^+$ 
and the same amount
of supersymmetries as our supergravity solution. 
We remark that the work of \cite{Blau} considered super Yang-Mills theory
on a generic curved spacetime 
with Killing spinors.
Here we have a specific choice of the metric but
our action is more general in that
we will allow for a time-dependent Yang-Mills coupling and
also we will introduce 
a non-topological axion 
term to the action. 
Both are necessary since  our SUGRA solution 
has a nontrivial dilaton and axion background.

\subsection{Construction}

The ${\cal N} = 4$ super Yang-Mills theory can be understood
as the dimensional reduction of the 10 dimensional super Yang-Mills theory.
Its action is
\be
S = \frac{1}{g_{YM}^2}\int d^4 x \mbox{Tr}\left(
-\frac{1}{4} F_{MN}F^{MN} - \frac{1}{2}
\bar{\Psi}\Gamma^{M}[D_{M}, \Psi]
\right),
\label{10Daction}
\ee
where $\Psi$ is a 10D Majorana-Weyl spinor and
\be \label{FMN}
F_{MN} = i[ D_{M}, D_{N} ], \qquad
D_{M} = \left\{ \begin{array}{ll}
\del_\mu - i A_\mu & (\mu = 0,1,2,3), \\
-i A_a & (a = 4, \cdots, 9).
\end{array} \right.
\ee
The action is invariant under the supersymmetry (SUSY)
transformation
\be
\d A_{M} = \frac{1}{2} \bar{\epsilon} \Gamma_{M} \Psi, \qquad
\d \Psi = - \frac{1}{4} F_{MN} \Gamma^{MN} \epsilon
\label{susytransf1}
\ee
for both $\epsilon = \eta$ (Poincare SUSY) and 
$\epsilon = x^\mu \Gamma_\mu \eta$ (conformal SUSY),
where $\eta$ is an arbitrary constant Majorana-Weyl spinor, and
$\Gamma^{MN} = \frac{1}{2}[\Gamma^{M}, \Gamma^{N}]$.
We use the convention that $\eta_{MN} = \mbox{diag}(-1, 1, \cdots, 1)$.

For compassion with our supergravity solution, 
we can also go to the lightcone coordinate.  Define
\be
\Gamma_{\pm} = \frac{1}{2}\left(\Gamma_0 \pm \Gamma_1\right)
\ee
so that
\be \label{ga2}
\{\Gamma_+, \Gamma_-\} = -1.
\ee
The Minkowski metric has
$\eta_{+-} = \eta_{-+} = -1/2$.
Using \eq{ga2}, we note that  a generic fermion $\Psi$ 
can always be decomposed as
\be
\Psi = \Psi_{+-} + \Psi_{-+},
\ee
where
\be
\Psi_{+-} = - \Gamma_+ \Gamma_- \Psi, \qquad
\Psi_{-+} = - \Gamma_- \Gamma_+ \Psi.
\ee
In view of the supersymmetry preserved by our supergravity solution,
we look for supersymmetric Yang-Mills theories
(SYM) preserving the supersymmetry
for $\epsilon$ being a constant 10D Majorana-Weyl spinor satisfying
\be
\Gamma_- \epsilon = 0.
\label{g-eps}
\ee
In this case, the SUSY transformation parameter satisfies
\be
\Gamma_- \Gamma_+ \epsilon = - \epsilon.
\ee

Let us start by rewriting the original SYM action (\ref{10Daction})
for a curved base space:
\be
S_0 = \int d^4 x ( {\cal L}_{0B} + {\cal L}_{0F} ),
\label{SC}
\ee
where
\bea
{\cal L}_{0B} &=& f_B 
% \frac{1}{g_{YM}^2}
\mbox{Tr} \left[
-\frac{1}{4} \tilde{g}^{MM'}\tilde{g}^{NN'}
F_{MN} F_{M'N'} \right], \label{L0B}\\
{\cal L}_{0F} &=& f_F 
% \frac{1}{g_{YM}^2}
\mbox{Tr} \left[ - \frac{1}{2} \bar{\Psi} \gamma^M [ D_M, \Psi] \right].
\label{L0F}
\eea
Since the base space is curved,
the covariant derivative $D_M$ also includes
the spin connection when it acts on $\Psi$.
The functions $f_B$ and $f_F$ are assumed to be functions
of $x^+$ only.
We will match $f_B$ with the factor $\frac{\sqrt{-g}}{g_{YM}^2}$,
where both the measure $\sqrt{-g}$ and
Yang-Mills coupling $g_{YM}$ are time-dependent functions.
Naturally we also deform the SUSY transformation
\eq{susytransf1} to include
two functions $f_A(x^+)$ and $f_{\Psi}(x^+)$
\be
\d A_{M} = \frac{1}{2} f_A \bar{\epsilon} \gamma_{M} \Psi, \qquad
\d \Psi = - \frac{1}{4} f_{\Psi} F_{MN} \gamma^{MN} \epsilon
\label{susytransf2}
\ee
For the parameter $\epsilon$ satisfying
(\ref{g-eps}),
these can  be written in more detail as
\bea
\d A_- &=& 0, \label{dA-} \\
\d A_+ &=& \frac{1}{2} f_A \bar{\epsilon} \gamma_+ \Psi_{-+}, \\
\d A_m &=& \frac{1}{2} f_A \bar{\epsilon} \gamma_m \Psi_{+-} \quad
\mbox{for} \quad m = 2, 3, \cdots, 9, \\
\d \Psi_{+-} &=& - f_{\Psi} F_{-m} \gamma^m \gamma_+ \epsilon, \\
\d \Psi_{-+} &=& f_{\Psi} \left( F_{-+} - \frac{1}{4} \sum_{m, n = 2}^9
F_{mn} \gamma^{mn} \right) \epsilon. \label{dPsi-+}
\eea
We will use indices $M, N = +, -, 2, 3, \cdots, 9$,
and indices $m, n = 2, 3, \cdots, 9$.
Below we will also use $i, j = 2, 3$ and
$a, b = 4, 5, \cdots, 9$.

Choose the metric of the base space to be of the form
\be \label{YM-metric}
ds^2 = \tilde{g}_{\mu\nu} dx^{\mu} dx^{\nu} =
- \tilde{k}^2 dx^+ dx^- + \tilde{h} dx^+ dx^+ + dx^i dx^i.
\ee
We remark that, here we use $\gt_{\m\n}, \tilde{k}, \htt$ to denote 
quantities in the Yang-Mills theory, in order to distinguish
them from the SUGRA quantities: $g_{\m\n}, k, h$.
For the vielbein, we take 
\be
E^- = \tilde{k}^2 dx^- - \tilde{h} dx^+, \qquad
E^M = dx^M \quad \mbox{for} \quad M \neq - ,
\label{spinconn1}
\ee
where we have extended the definition of the vielbein
to the indices $a = 4, \cdots, 9$,
so that we can define
\be
\gamma_M = E_M{}^A \Gamma_A.
\ee
We have $\{ \gamma_{M}, \gamma_{N} \} = 2\tilde{g}_{MN}$.
A different choice of vielbein is possible and corresponds to
a different choice of the spin connection.
For our choice, the spin connection can be easily read off
from (\ref{spin}) by setting $r = dr = 0$.

One can check that the variation of the action \eq{SC}
for a generic base space is
\bea
\d S_0 = \int d^4 x \; \mbox{Tr} &\Big[ &
- \frac{1}{2} \bar{\epsilon} 
\Big(
\frac{1}{4} f_{\Psi} \gamma_{MN}
[D_{K}, f_F \gamma^{K}] \Psi +
f_B [ D_{M}, f_A \gamma_{N} ]\Psi
\nn \\
&& 
+(f_F f_{\Psi} - f_B f_A)\gamma_M [D_N, \Psi]
\Big) F^{MN}
\Big].
\eea
For our choice of the vielbein \eq{spinconn1},
both terms vanish for $\epsilon$ satisfying \eq{g-eps} if
\be
f_B f_A - f_F f_{\Psi} = 0, \qquad
f_B f'_A - f'_F f_{\Psi} = 0.
\ee
When the coupling is constant \cite{Blau},
we can choose $f_B = f_F = \sqrt{-g}/g_{YM}^2$
and $f_A = f_{\Psi} = \sqrt{-g}$.
In our case the coupling $g_{YM}$ is not a constant,
by convention we identify $f_B$ with $\sqrt{-g}/g_{YM}^2$.
(This is how one defines $g_{YM}^2$).
Note that one can scale $\Psi$ by an arbitrary function
of $x^+$.
In particular we can scale $\Psi$ such that
$f_F$ is equal to $f_B$.
This implies that $f_A$ equals $f_{\Psi}$
and so the solution is
\be
f_B = f_F = f_A = f_{\Psi} = \frac{\sqrt{-g}}{g_{YM}^2}
\ee
up to scaling $f_A$ and $f_{\Psi}$ by a constant,
which is equivalent to scaling $\epsilon$.
We see that the super Yang-Mills theory can
have a generic coupling function $g_{YM}$ depending on $x^+$.

There is an additional term invariant under
the same SUSY transformation
\bea
{\cal L}_{\chi} &=& \chit(x^+) 
\mbox{Tr} \left( \frac{1}{4} \epsilon^{\m\n\r\s} F_{\m\n}
F_{\r\s}
+ \frac{1}{2} \frac{f_A}{f_{\Psi}}
\bar{\Psi}_{+-}\Gamma^2\Gamma^3\Gamma_-\Psi_{+-} \right),
\label{L2}
\eea
where $\epsilon^{\mu\nu\alpha\beta}$ is
the totally antisymmetrized tensor
with $\epsilon^{+-23} = 1$.
It is not hard to show that under the variations \eq{dA-} - \eq{dPsi-+},
${\cal L}_2$ is invariant for an arbitrary function $\chit(x^+)$.
In conclusion, a super Yang-Mills Lagrangian invariant under the transformation
(\ref{dA-})-(\ref{dPsi-+}) is
\be
{\cal L} = {\cal L}_{0B} + {\cal L}_{0F} + {\cal L}_{\chi}.
\ee
Explicitly, the total action  is 
\bea
S &=& \int d^4 x \;
\frac{\sqrt{-g}}{g_{YM}^2}\mbox{Tr}
\Bigg[ 2 \tilde{k}^{-4} F_{+-}^2
+ 2 \tilde{k}^{-2} F_{+i} F_{-i} - \frac{1}{2} F_{23}^2   \nn\\
&& + 2 \tilde{k}^{-2} [ D_-, X_a ] [ D_+, X_a ] 
- \frac{1}{2} [D_i, X_a]^2 + \frac{1}{4} [X_a, X_b]^2
 \nn \\
&&  -4 \tilde{k}^{-4} \htt(x^+, x^2, x^3)
\left( \frac{1}{2} F_{-i}^2 + \frac{1}{2} [D_-, X_a]^2
\right) \Bigg]\nn \\
&+&  \int d^4 x 
\frac{\sqrt{-g}}{g_{YM}^2}
\mbox{Tr} \Bigg[
\bar{\Psi}_{-+} \Gamma_+ [D_-, \Psi_{-+}]
+ \bar{\Psi}_{+-} \tilde{k}^{2} \Gamma_- [D_+, \Psi_{+-}]   
- \frac{1}{2} \bar{\Psi} \Gamma^i [D_i, \Psi]
+ \frac{i}{2} \bar{\Psi} \Gamma^a [X_a, \Psi]   \nn \\
&&   + \tilde{k}^{-4} \htt(x^+, x^2, x^3)
\bar{\Psi}_{+-} \tilde{k}^{2} \Gamma_- [D_-, \Psi_{+-}]
\Bigg]\nn \\
&+&  \int d^4 x \;
\mbox{Tr}\Bigg[
\chit(x^+) \left( \frac{1}{4} \epsilon^{\m\n\r\s} F_{\m\n} F_{\r\s}
+ \frac{1}{2} \bar{\Psi}_{+-}\Gamma^2\Gamma^3\Gamma_-\Psi_{+-}
\right) \Bigg]. 
\label{SYMaction}
\eea
In the above, we have denoted $A_a$ as $X_a$
and $i = 2, 3$, $a = 4, \cdots, 9$.
This action is invariant under the 8 supersymmetries of
\eq{dA-} - \eq{dPsi-+}.
One can verify that the conformal supersymmetry are all broken.
Thus our theory preserves 8 supersymmetries.

The above is for general $\htt$.
We note that if the function $\htt$ is bilinear in $x^2, x^3$:
\be \label{h-sym}
\htt(x^+, x^2, x^3) =  \htt_{ij} (x^+) x^i x^j ,
\ee
then our action (\ref{SYMaction}) enjoys the scaling symmetry:
\bea 
&x^+ \rightarrow x^+, \qquad
x^- \rightarrow \lambda^2 x^-, \qquad
x^i \rightarrow \lambda x^i, \label{scaling-sym} \\
&\Gamma_+ \rightarrow \lambda \Gamma_+, \qquad
\Gamma_- \rightarrow \lambda^{-1} \Gamma_-, \qquad
\Gamma_i \rightarrow \Gamma_i, \qquad
\Gamma_a \rightarrow \Gamma_a, \\
&A_+ \rightarrow A_+, \quad
A_- \rightarrow \lambda^{-2} A_-, \quad
A_i \rightarrow \lambda^{-1} A_i, \quad
X_a \rightarrow \lambda^{-1} X_a, \quad
\Psi \rightarrow \lambda^{-3/2} \Psi, \nn
\eea
since $\tilde{h}$ scales like
\be
\htt(x^+, x^2, x^3) \rightarrow \lambda^2 \htt(x^+, x^2, x^3)
\ee 
in this case.
This symmetry is also a symmetry
of the original AdS/CFT background since 
it can be viewed as a combination of the usual
scaling symmetry
\be  \label{full-s}
x^{\mu} \rightarrow \lambda x^{\mu}
\ee
and the  Lorentz boost in the $x^1$ direction
\be
x^+ \rightarrow \lambda^{-1} x^+, \quad
x^- \rightarrow \lambda x^-, \quad
x^i \rightarrow x^i.
\ee
However the full scaling symmetry \eq{full-s} is broken in our case and only 
the partial scaling symmetry is respected by our solution.

\subsection{Born-Infeld analysis} \label{DBI}

In this section, we propose, and give further
justification, that 
the time-dependent SYM theory we constructed 
in section 3.1 is dual to the 
the string theory based on the time-dependent 
\IIB background we constructed in section 2.

Our proposal is that the time-dependent SYM theory
\eq{SC} with Yang-Mills coupling given by
\be \label{iden1}
g_{YM}^2 = g_s \equiv e^{\phi} 
%\equiv \frac{\tilde{k}^2}{2}.
\ee
and $\htt$ given by \eq{h-sym} 
provides a dual description of the string theory 
based on the time-dependent \IIB background
\eq{ansatz-metric}-\eq{ansatz-scalar} with $h$ given by
\eq{h-sugra}. Moreover we propose the following identification:
\be  \label{iden2}
h = \htt, 
\qquad  \chi = \chit, \qquad k = \tilde{k}.
\ee
We remind the reader again that the
left hand side are SUGRA quantities and
the right hand side are SYM quantities.

Let us now explain and justify our proposal. 
Consider a single D3-brane in our
supergravity background.  The D3-brane action is given by the DBI action
plus a coupling to the RR gauge fields
\be
S = - \mu_3 \int d^4 x e^{-\phi} \left[
- \det \left( G_{\m\n} + {\cal F}_{\m\n} \right) \right]^{1/2}
+ \int C \wedge e^{{\cal F}},
\ee
where
\be
{\cal F}_{\m\n} \equiv B_{\m\n} + 2\pi\alpha' F_{\m\n},
\ee
\be
C = C^{(0)} + C^{(2)} + \cdots, 
\ee
are the RR gauge fields, and  
\be
G_{\m\n} =  \frac{\del X^M}{\del x^\m}
\frac{\del X^N}{\del x^\n} g^{(s)}_{MN}, \quad 
B_{\m\n} =\frac{\del X^M}{\del x^\m}
\frac{\del X^N}{\del x^\n} B_{MN} 
\ee
are the pull back to  D3-brane worldvolume of the 
spacetime metric in the string frame and the NSNS B-field. 

By performing a weak field expansion, we have
\be \label{weak}
S_{DBI}= -\m_3(2\pi \a')^2 \int d^4 x e^{-\phi} \sqrt{-\det(G_{\m\n})}
\left(
-\frac{1}{4} F_{\m\m'} F_{\n\n'} G^{\m\n} G^{\m'\n'}
+ \cdots\right),
\ee
where $\cdots$ denotes higher order terms in $F G^{-1}$. 
Now let us consider a D3-brane placed at $u=u_0 $ and extends in the $+,-,2,3$
directions. Take a static gauge
\be
x^\m= X^\m, \quad \m = +,-, 2,3.
\ee
We have 
\be
G_{\m\n} = \frac{R^2}{u_0^2}e^{\phi/2} \hat{g}_{\m\n},
\ee
where
\be
\gh_{\m\n} = \begin{pmatrix}
h & -k^2/2 &0&0 \\
-k^2/2 &0&0&0\\
0&0&1&0\\
0&0&0&1
\end{pmatrix}.
\ee
Substituting into \eq{weak}, we have 
\be \label{weak1}
S_{DBI}= -\m_3(2\pi \a')^2\int d^4 x \left[
-\frac{ 1}{4} \sqrt{\hat{g}} e^{-\phi}F_{\m\m'} 
F_{\n\n'} \gh^{\m\n} \gh^{\m'\n'} \right]
+ \cdots.
\ee
This is precisely the form of the SYM action \eq{SC}
by identifying the string coupling with the Yang-Mills coupling
in the usual way
\be
e^{\phi} = g_{YM}^2,
\ee
and identifying $\gt_{\m\n}$ of \eq{YM-metric} with $\gh_{\m\n}$
here. The later implies that 
\be
\tilde{k} = k, \quad \htt = h .
\ee
Moreover since in our case $B=0$ and there is only
the $C^{(0)}$ RR field, 
the RR coupling reduces to $\int C^{(0)} F \wedge F$
and can be identified with the SYM piece $\int \cL_{\chi}$ directly.
Therefore we find that the weak field expansion
of the bosonic D3-brane action produces precisely
our time-dependent SYM theory. We remark that the
higher order terms in \eq{weak1} vanish as we take $u_0 \to 0$.

This justifies our choice \eq{iden1} and our
identification \eq{iden2} for the functions 
which appear in our supergravity
solution and in our SYM Lagrangian.
For $N$ D3-branes the action is
given by the nonabelian generalization of \eq{weak1}.
Our action \eq{SYMaction} is the supersymmetric
completion of it.

For the duality to be precise, we still need to determine the radius $R$
of the SUGRA solution in terms of gauge theory parameters. Recall that
our SUGRA solution can be obtained as a near horizon limit of the SUGRA
solution for a stack of D3-branes with pp-wave on it. Consider a stack
of $N$ such D3-branes. If one equates the mass and charge of the
D-brane, one obtain that
\be \label{RR}
R^4 = 16 \pi N \langle g_s^{-1} \rangle^{-1} l_s^4
\ee
where
$\langle g_s^{-1} \rangle := \int dx^+ \tilde{k}^2 e^{-\phi} /
\int dx^+ \tilde{k}^2$
is the $x^+$-average of the inverse of the string coupling
$g_s = e^{\phi}$.

Provided that $\langle g_s^{-1}\rangle$ is well defined, we propose that
the time-dependent SYM theory is dual to the quantum gravity in the bulk
with $R$ given by \eq{RR}. For instance, $\langle g_s^{-1}\rangle$ is
well defined for the example \eq{eg-2}. Our proposal is supported by a
number of matchings. First we see that there is a precise matching
between the functional dependence on $x^+$ of the two theories.
We will also explain in sec.\ref{Einstein}
how the Einstein equation (\ref{Ein}) is
realized as a constraint in the SYM theory.
Furthermore, our theories also
match in their various symmetries.
The SYM action \eq{SC} enjoys a global $SO(6)$
invariance rotating the six scalars.
This is mapped to the rotational symmetry of
the $S^5$ on the supergravity side.
Supersymmetry also matches.
Both theories observe 8 supersymmetries, and these unbroken
supersymmetries satisfy the same chirality condition $\C^+ \e =0$.
Moreover, as we noted above, both the supergravity solution and the SYM
action observe a scaling symmetry \eq{scaling-sugra} and
\eq{scaling-sym}.

While the matching between the weak field expansion
of the DBI action with the super Yang-Mills theory
only makes sense in the low energy limit,
it is possible that the identification of parameters
(\ref{iden1}) and (\ref{iden2}) between
the type \IIB string theory and super Yang-Mills theory
could be modified by higher derivative terms
in the $\alpha'$ expansion.
The matching of parameters including higher order terms
can in principle be achieved order by order by comparing
the type \IIB stringy corrections to the supergravity
equations of motion with the quantum corrections
of super Yang-Mills theory via a similar calculation
as the one carried out in sec. \ref{Einstein}
but to a higher order.

\section{Holographic duality}

\subsection{Two-point correlation functions and singularity structures}

In AdS/CFT duality, the action of fluctuations in AdS space with
specified boundary conditions is matched with
correlation functions of the corresponding operators.
As a result, the boundary bulk propagator in AdS
should agree with the corresponding two-point correlation functions.

\underline{SUGRA calculation}

In this subsection, we will write down the metric (\ref{ansatz-metric})
in the Rosen form (\ref{metric-diag})
\begin{equation}
ds^2 = \frac{1}{u^2} (du^2 - 2 dx^+ dx^- + a_{ij}(x^+) dx^i dx^j )
= g_{\mu\nu}dx^{\mu}dx^{\nu}.
\end{equation}
The $S^5$ part of the spacetime will be ignored in this section
for simplicity.
In our case we have $d = 4$, but for generality
we leave $d$ as a variable in the following.
Consider the action of a scalar field $\varphi$
\be
S= \frac{1}{2} \int du d^d x \sqrt{-g} \, (g^{MN} \del_M \varphi \del_N
\varphi + m^2 \varphi^2).
\ee
First we want to solve the equation of motion
\be
\Box \varphi - m^2 \varphi = 0. 
\ee
One has 
$\Box = u^{d+1} \del_u (u^{-d+1} \del_u) + u^2 \Delta$ where
\be
\Delta:=\frac{1}{\sqrt{a}}
\Big[ - \del_+ (\sqrt{a} \del_-) - \del_- (\sqrt{a} \del_+) 
+ \del_i(\sqrt{a}a^{ij} \del_j)\Big]
\ee
is the $d$ dimensional d'Alembertian operator of the boundary metric
and we have denoted 
the determinant of the matrix $a_{ij}$ by
$ a$. 
Take the ansatz
\be
\varphi = \vphit(\vk,u) \psi_\vk(\vx)
\ee
and use the technique of separation of variables, 
we find 
\be \label{eq1}
u^{d-1} \del_u (u^{-d+1} \del_u \vphit) -(k^2 +\frac{m^2}{u^2}) \vphit
=0
\ee
and
\be \label{eq2}
\Delta\psi_\vk = -k^2 \psi_\vk
\ee
for arbitrary separation constant  $k^2 \in \bR$
\footnote{Note that here $k^2$ is simply a constant of separation. One may 
also introduce the inner product 
$k\cdot k := a_{ij}(x^+) k^i k^j -k_+ k_-$ for some momentum vector $k$.
This object however will not be used at all in this paper.}.
The equation \eq{eq2} has solution
\be
\psi_\vk(\vx) = \frac{1}{a^{1/4}} e^{i(k_i x^i - k_- x^- - \beta(x^+))}, 
\ee
where $\vk$ denotes the set $(k_i,k_-,k^2)$,
\begin{equation}
\dot{\beta} = \frac{a^{ij}(x^+) k_i k_j - k^2}{2k_-}
\end{equation}
and $a^{ij}$ is the inverse matrix of $a_{ij}$.
We note that $\psi_k$'s form a basis of functions of $x$.
One can check that
\be \label{comp1}
\int d^d \vx \sqrt{a} \psi_k(\vx) \psi_{k'}(\vx)
= 2 k_- \delta(k_i + k'_i) \delta(k_- + k'_-) \delta(k^2 - k'{}^2).
\ee
We also have
\be \label{comp2}
\int [d^d \vk] \psi_k(\vx) \psi^{\ast}_k(\vy) = \frac{1}{\sqrt{a}}
\d(x^+ - y^+)\d(x^- - y^-)\prod_i \d(x^i - y^i).
\ee
where the measure of integration is
\be
\int [d^d \vk] := 
\int^\infty_{-\infty} d^{d-2}k_i  
\int^\infty_{-\infty}\frac{d k_-}{2k_-} 
\int^\infty_{-\infty}d(k^2) .
\ee

The equation \eq{eq1} depends on the separation constant $k^2$.
Its most general solution which is asymptotic to
$\e^{2h_-} \vphit_0(k^2)$ is
\be \label{gen-bb}
\vphit(k^2,u) = K_\e(k^2,u) \vphit_0(k^2),
\ee
where 
\be \label{bb-K}
K_\e(k^2,u) = \frac{\vphit^{(-)}(k^2,u)+ A(k^2)\vphit^{(+)}(k^2,u) }
{\vphit^{(-)}(k^2,\e)+A(k^2)\vphit^{(+)}(k^2,\e)} \; \e^{2h_-}
\ee
is the bulk-boundary Green function. 
Here $h_\pm = (d \pm 2 \nu)/ 4$ and
$\nu := \frac{1}{2}\sqrt{d^2+4m^2} > 0$.
$\vphit^{(-)}$ is a non-normalizable solution 
which behaves as $u^{2h_-}$ as $u \to 0$, and $\vphit^{(+)}$ is a
normalizable solution which behaves as $u^{2h_+}$ as $u \to
0$\footnote{
Explicitly, for $k^2 <0$, we have 
\be \label{vphit}
\vphit^{(\pm)} (k^2,u) \propto u^{d/2} J_{\pm\nu}(|k|u).
\ee
if $\nu$ is non-integral. If $\nu$ is integral, the two independent
solutions are $\vphit^{(+)}$  in \eq{vphit} and
\be
\vphit^{(-)} (k^2,u) \propto u^{d/2} Y_{\nu}(|k|u).
\ee
Here $|k|= +\sqrt{|k^2|}$.
}.   
$A(k^2)$ is an arbitrary coefficient which 
corresponds to the freedom to specify the vacuum state of the dual 
field theory, which is reflected in  the freedom to choose  different
Lorentzian propagator in the dual theory.

Expanding the scalar field in terms of this basis
\be
\varphi(x, u) = \int [d^d \vk] \psi_\vk(x) \vphit(k^2, u),
\ee
and plugging it into the action,
we find
\be \label{Sphi}
S = - \lim_{u = \epsilon\rightarrow 0} \int [d^d \vk]
\Big( u^{-d+1} \vphit(k^2,u) \del_u \vphit(k^2,u) \Big).
\ee
Now consider an operator $\tilde{\cO}$ on the boundary theory 
which couples to the field $\vphit_0$ 
with the coupling $\int [dk] \tilde{\cO}(\vk)\vphit_0(\vk)$.
This gives
\be
\langle \tilde{\cO}(\vk) \tilde{\cO}(\vk') \rangle = 
\frac{\d^2 S}{\d \vphit_0(\vk) \d \vphit_0(\vk')}.
\ee
Substitute  the  general solution \eq{gen-bb}
in (\ref{Sphi}),
we obtain 
\be
\langle \tilde{\cO}(\vk) \tilde{\cO}(\vk') \rangle 
= - \e^{-d+1} \d(k_i+k_i')
\d(k_-+k_-') \d(k^2 -k^2{}') 2 k_- \lim_{u=\e \to 0} \del_u K(k^2,u).
\ee
The resulting two-point function depends strongly on the coefficient
$A(k^2)$. 
We note that the 
nontriviality of the metric affects only the
eigenfunction $\psi_\vk(\vx)$,
but the modes $\vphit^{(\pm)}$ as well as the 
bulk-boundary propagator
$K_\e(k^2,u)$ as given in \eq{bb-K} take exactly the same form as in the
standard AdS case. 
Thus one can imagine deforming adibatically the metric back to the undeformed AdS metric and 
%cc2 we should have 
use the same
$K_\e(k^2,u)$. In that case
$A(k^2)$ is chosen
via analytic continuation from the Euclidean  and it is 
\be
\lim_{u=\e \to 0} \del_u K_\e(k^2,u) = \e^{2 \nu -1} k^{2 \nu} + \cdots.
\ee
Here $\cdots$ denotes terms that are of sub-leading order in $\e$ and
terms which contain integral powers in $k^2$, which as usual give rise to 
contact terms in the correlation function and thus will be ignored.
To go to the coordinate space, we consider the operators
\bea 
&& \hat{\cO}(\vx) := a^{\frac{1}{4}}(x^+)
\int [d^d \vk] \psi_\vk(\vx)\tilde{\cO}(\vk),\label{def-O} \\
&& \hat{\varphi}_0(\vx) :=  a^{\frac{1}{4}}(x^+)
\int [d^d \vk] \psi_\vk(x) \vphit_0(k^2).\label{def-phihat}
\eea
Note that, different from the usual Fourier transform, 
we have included an additional factor of  $a^{\frac{1}{4}}(x^+)$ in
front. This is a natural definition in view of 
the identity
\be
\int [dk] \cO(\vk) \vphit_0(\vk)= \int dx  \hat{\cO}(\vx) \hat{\varphi}_0(\vx).
\ee 
Also our results will turn out to be simpler when expressed in terms of
$\hat{\cO}$. Using 
\be
\langle \hat{\cO}(\vx) \hat{\cO}(\vy) \rangle = 
a^{\frac{1}{4}}(x^+)a^{\frac{1}{4}}(y^+)
\int [d \vk] [d \vk']
\psi_\vk(\vx) \psi_{\vk'}(\vx')
\langle \tilde{\cO}(\vk) \tilde{\cO}(\vk') \rangle,
\ee
it is easy to obtain
\be \label{2pt-sugra}
\langle \hat{\cO}(\vx) \hat{\cO}(\vy) \rangle = 
%cc2 sign corrected
\frac{c}{(B_{ij} \d x^i \d x^j- \d x^-  )^{ \frac{d}{2}+ \nu}} 
\frac{1}{(\det B^{ij})^{\frac{1}{2}} (\d x^+)^{\nu+1}}, 
\ee
where $c$ is a constant. Here $\d x^{i,\pm}:= x^{i,\pm} -y^{i,\pm}$, 
\be B^{ij} := \int_{x^+}^{y^+} dx^+a^{ij} \ee
and $B_{ij}$ is the inverse.
If $a_{ij}$ is proportional to the unit matrix as in our example \eq{eg-1}
or \eq{eg-2}, 
\be \label{a-unit} 
a_{ij} = M^2 \d_{ij},
\ee
then
\be
B^{ij} = \d_{ij} (f(x^+) - f(y^+)):= \d_{ij} \cdot \d f, \quad \mbox{where} 
\quad f(x^+) = \int^{x^+} \frac{dz^+}{M^2(z^+)}
\ee
and
\be \label{2pt-sugra2}
\langle \hat{\cO}(\vx) \hat{\cO}(\vy) \rangle = 
%cc2
\frac{c}{( (\d x^i)^2 - \d x^- \d f)^{ \frac{d}{2}+ \nu}} 
\left(\frac{\d f}{\d x^+}\right)^{\nu+1}.
\ee
This expression reduces to the usual result for flat
space when $M^2 =1$. 
For the example \eq{eg-1} or \eq{eg-2},
we have $M^2 = (x^+)^\a$
and
\be
f(x^+) = \frac{(x^+)^{1-\a}}{1-\a}.
\ee
Notice that \eq{2pt-sugra2} is singular if both $x^+$ and $y^+$
approach zero simultaneously ($0<\alpha<1$).

We remark  that 
in general for an operator $\cO$ with scaling dimension $\D$, i.e.  under 
the transformation \eq{scaling-sym}, $\cO$ transforms as
$ \cO(x') = \l^{-\D} \cO(x)$,  
the most general form of the two-point function that is compatible 
with the symmetry of the theory is:
\be \label{2pt-kinematic}
\langle \cO(\vx) \cO(\vy) \rangle = |\d x^-|^\D \; 
g\Big(x^+,y^+,\frac{\d x^i}
{\sqrt{|\d x^-|}}\Big),
\ee
where $g$ is an arbitrary function. 
Our result \eq{2pt-sugra} as determined by the bulk-boundary propagator 
approach is compatible with this for  $\D = \frac{d}{2} +\n$. 
However \eq{2pt-sugra} is more specific than the kinematical result
\eq{2pt-kinematic}. 
It is a consequence of dynamics.
  
\underline{Field theory calculation}

Next we compute the 
two-point correlation function from the field theory
point of view. Consider a scalar field $\varphi$ which
satisfies the equation of motion
\be
(\D -m^2) \varphi =0.
\ee
It has the mode expansion
\be \label{129}
\varphi = 
%cc2 
\int_0^\infty \frac{d k_- }{\sqrt{2 k_-}}\int^\infty_\infty dk_i  
\; (\psi_\vk a_\vk 
+ \psi_\vk^* a_\vk^\dagger ),
\ee
where $\psi_\vk$ is given in \eq{eq2} above
%cc2
and $k^2 = -m^2$.
The equal time commutation relation
\be
[ \varphi(\vx), \sqrt{a}\del_-\varphi(\vy)]\Big|_{x^+ = y^+} = i \d(x^- -y^-)
\d(x^i - y^i)
\ee
implies that
\be
[a_\vk, a_{\vk'}^\dag] = \d(k_- - k_-') \d (k_i -k_i') .
\ee
Using this, one can easily compute the time ordered (with respect to $x^+$)
product $\langle  T \varphi(\vx) \varphi(\vy)  \rangle $. 
We have
\be
G(\vx,\vy) := -i \langle  T \varphi(\vx) \varphi(\vy) \rangle
=  \int [d^d \vk]  \; \frac{1}{k^2 +m^2 
- i \e}\; \psi_\vk (x) \psi_\vk (y)^* .
\ee
Here the vacuum is choosen to be anhillated by the operators $a_\vk$.
This corresponds to the choice of the bulk-boundary Green function
$K_\e$ above in the SUGRA calculation. 
For the interests in studying singular spacetime,
let us consider the example 
\eq{eg-2} of the metric.
%p7
%In general for a metric defined on
%a half interval $0 < x^+ < \infty$, we also need to
%specify the boundary condition at $x^+=0$ to be either Neumann or
%Dirichlet. But in case where the volume factor $\sqrt{-g} \to 0$ (as in
%\eq{a-unit}), any $\varphi$ is allowed so long as $\varphi$ is nonsingular
%at $x^+=0$. In this case we can continue to use the mode expansion \eq{129}
%above.
The Green function $G$ can be easily be computed and the result is 
%cc2
(for $m^2=0$)
\be \label{2pt-field}
%cc2
G(\vx,\vy) = \frac{c}{( (\d x^i)^2- \d x^- \d f)^{\frac{d}{2}-1}} 
\frac{1}{a^{\frac{1}{4}}(x^+)a^{\frac{1}{4}}(y^+)},
\ee 
The additional (factorisable) factors of $a$ suggests one to consider
the rescaled field $\hat{\varphi} := a^{\frac{1}{4}}\varphi$. Note
that this is
the same rescaling appearing  above in the definition of the dual
field operators \eq{def-phihat}.

Now we want to compare our SUGRA result and the field theory result.
In the usual case, the form of the two-point
function is fixed by the conformal symmetry.
In our case, the scaling symmetry does not fix the form of the 
two-point function uniquely.
So generally there is no reason to expect the two
computations to agree.
In fact for an operator 
$\hat{\cO} := (\hat{\varphi})^n$, the two-point function is
\be \label{2pt-field2}
\langle \hat{\cO}(\vx) \hat{\cO}(\vy) \rangle =
(a^{\frac{1}{4}}(x^+)a^{\frac{1}{4}}(y^+) G(\vx,\vy))^n  = 
%cc2 change C to c
\frac{c}{\big( (\d x^i)^2- \d x^- \d f\big)^{ n(\frac{d}{2}-1)}} 
\ee
in the tree level approximation. 
This is to be compared with \eq{2pt-sugra2}. The factor involving $\d
f$ fixes $\D=n(\frac{d}{2}-1)$. 
However due to the absence of
the term $(\d f /\d x^+)^{\nu +1}$ in \eq{2pt-field2},
\eq{2pt-sugra2} and \eq{2pt-field2} cannot agree with each other .
In particular the singularity structure is different. 
It is remarkable that \eq{2pt-field}
is completely regular even when $x^+, y^+ \to 0$, 
while the SUGRA result \eq{2pt-sugra2} is singular.
%p8
Note that there is no particle creation in
either the bulk or boundary theory.
This can be easily checked following
similar computations as \cite{gibbons,Vijay}.
Hence there is no ambiguity in
the two-point functions associated with
the choice of vacuum.

Our interpretation of the result is the following: The SUGRA is a low
energy approximation. The singularity at $x^+=0$ of the spacetime as
revealed by the divergence in $R_{++}$ and by the two-point function
\eq{2pt-sugra2} is just a low energy description which may be
modified in the full quantum gravity by string loop and $\a'$
effects. 
As we proposed, the  quantum theory
is described in terms of the dual quantum SYM we constructed in section
3. On the SYM side, the singularity structure of the spacetime, as
revealed by the two-point function, is indeed different. 
Although the field theory result is computed at weak coupling, there
is reason to expect that it will be able to capture the 
qualitative behaviour of the spacetime singularity \cite{shenker}.
%p7
(If we choose the plus sign for $\phi$ in (41),
the string coupling actually goes to zero at
the singularity.)
Therefore our result suggests that the singularity is resolved
to some extent. 
Although our SYM computation is preliminary
as we have done it only at the free and tree level,
we believe that the picture and interpretation is basically correct.
What is surprising is that it seems that
$\a'$ effects, rather than string
loop effects,
are sufficient to smoothen 
the spacetime singularity. More work is needed on this issue.

\subsection{Einstein equation from super Yang-Mills theory}
\label{Einstein}

On the supergravity side, Einstein's equation imposes a constraint
\eq{cond-red} among the parameterising functions
\begin{equation}
\frac{1}{2}(\phi')^2+\frac{1}{2}e^{2\phi}(\chi')^2 = -h_{22}-h_{33} .
\label{Ein}
\end{equation}
The proposed duality implies that this constraint should also be
imposed on the super Yang-Mills theory.
But why?
At the classical level, the super Yang-Mills theory is well defined
regardless of the Einstein equations.
However, since the duality mixes classical effects and quantum effects
between the dual theories, a super Yang-Mills theory is a candidate
of the dual theory only if it is well defined at the quantum level.
One should demand that all correlation functions
of fundamental operators are well defined,
and that the scaling symmetry is anomaly-free.

More specifically, we suggest that Einstein equations are
obtained from the super Yang-Mills theory by demanding
the vacuum expectation value (VEV) of the energy-momentum
operator $T_{\mu\nu}$ to be finite.
Assuming the duality, $\langle T_{\mu\nu}\rangle$
can be computed on the supergravity side \cite{deHaro}
\begin{eqnarray}
\langle T_{\mu\nu} \rangle = -\frac{1}{8\pi G_N}
\lim_{\epsilon\rightarrow 0} & \left[ \frac{1}{\epsilon^2}
\left( -g_{(2)}{}_{\mu\nu}+g_{(0)}{}_{\mu\nu} \mbox{Tr} g_{(2)}
+ \frac{1}{2} R^{(4)}_{\mu\nu} -\frac{1}{4} g_{(0)}{}_{\mu\nu} R^{(4)} \right)
\right. \nn \\
& \left. + \log \epsilon \left( -2h_{(4)}{}_{\mu\nu}-T^a_{\mu\nu} \right)
+ \cdots \right], \label{T}
\end{eqnarray}
where we only listed the diverging part of the VEV.
Let us explain the notation.
First, $\epsilon$ is the infrared cutoff at $u = \epsilon$.
The 4D functions $g_{(0)}$, $g_{(2)}$ and $h_{(4)}$
are expansion coefficients of the metric near
the AdS boundary
\be
g(x,u) = g_{(0)} + g_{(2)} u^2 + g_{(4)} u^4 + h_{(4)} u^4 \log u^2
+ {\cal O}(u^5),
\ee
while $g(x, u)$ is the 4D part of the full metric
of the 5D warped metric
\begin{equation}
ds^2 = \frac{R^2}{u^2}(du^2 + g_{\mu\nu}dx^{\mu}dx^{\nu}).
\end{equation}
$R_{\mu\nu}^{(4)}$ and $R^{(4)}$ are the Ricci tensor
and scalar curvature defined by the 4 dimensional metric $g_{(0)}$.

In order for $\langle T_{\mu\nu} \rangle$ to be finite,
both divergent terms should vanish.
For the background solutions in which the $u$ dependent terms
are suppressed, $g_{(2)}$ and $h_{(4)}$ vanish.
The logarithmic divergent term vanishes automatically.
The vanishing of the $1/\epsilon^2$ term implies
the 4 dimensional Einstein equation
\begin{equation}
R^{(4)}_{\mu\nu} - \frac{1}{2} g_{(0)}{}_{\mu\nu} R^{(4)} = 0
\end{equation}
(for $\mu, \nu = +, -, 2, 3$).
This is the source-free equation because
neither dilaton nor axion was included in (\ref{T}).
It is straightforward to repeat the computation of \cite{deHaro}
to include contribution of generic matter fields,
and, of course, the final result is to give the correct
Einstein equation with sources.

We remark that
the 4D Einstein equation for the boundary 
is in general not equivalent to the 5 dimensional
Einstein equation with a negative cosmological constant (flux).
However,
the 4D Einstein equation
happens to agree with the 5D Einstein equation
for a class of backgrounds including the ones under consideration.
In fact, only the 
$(++)$-component of the 4D Einstein tensor
is nontrivial $G_{++} = - h_{22} - h_{33}$.
For SUGRA backgrounds with generic 
$r$-dependence
in $g(r,x)$, the 5D SUGRA equations correspond to
the RG equations in the field theory \cite{dBVV}.

This argument is so far incomplete because
the energy momentum tensor computed above was
based on the validity of the duality. 
In general the duality may not
hold when the Einstein equation is not valid, in which the later
is precisely what we want to check. 
Thus we have to check independently that the VEV of the SYM
energy-momentum tensor is indeed given by (\ref{T}).
It has been argued in \cite{gibbons} 
that $\langle T_{\mu\nu}\rangle =0$ for a free theory. 
It is of course not the case for our time-dependent SYM. 
Consider a Yang-Mills theory coupled to fermions
living on a base space with a generic metric $\tilde{g}$.
The vacuum expectation value of the energy-momentum tensor
generically has UV divergences which need to be regularized
in a diffeomorphism-invariant way.
By a simple power counting,
one can see that $\langle T_{\mu\nu} \rangle$
has potentially a quartic, a quadratic and a logarithmic divergence.
By dimensional analysis,
the quartic divergence is a constant times
the cut-off (Planck) scale $M_P^4$.
This is just the cosmological constant
contribution from each quantum field.
It is well-known that it cancels for
a supersymmetric field theory.
The quadratic divergence must be of the form
\be
(a \tilde{R}_{\mu\nu} + b \tilde{g}_{\mu\nu}\tilde{R})M_P^2
\ee
with some numerical constant $a$, $b$.
The precise form may depend on the regularization scheme.
However, one should regularize it so that
this vacuum energy-momentum tensor is conserved.
This implies that it is proportional to
the 4D Einstein tensor, so that its effect can be absorbed
by renormalizing the 4D Newton constant.
The logarithmically divergent term
is not of interest for the purpose of this paper.
The finite part of $\langle T_{\mu\nu}\rangle$ has been
extensively studied in the context of conformal/Weyl anomaly
\cite{Weyl}.
A more detailed studies of these shall be interesting. 

\section{Discussion}

In this paper we have compared the two-point functions computed at
different regimes of the t'Hooft coupling. We  find that the two-point
function is not protected by nonrenormalization theorem and the SUGRA
result is different from the gauge theory result.  
Since our gauge theory is supersymmetric, it is plausable that there
will exist some modified form of nonrenormalization theorem. 
It is important to establish their existence and to use them to
compute quantities that are nonrenormalized. Such quantities will
allow one to compare field and SUGRA calculations directly and thus
provide
a check of the proposed gravity/gauge duality for the time
dependent background.

In the usual AdS/CFT correspondence, the scaling symmetry of the AdS
background implies that the dual gauge theory is conformal. Given that
the conformal symmetry is preserved at the quantum level, this conformal
symmetry has been a powerful tool for analysing the field theory and
providing valuable understanding
of the duality. In our case, 
let us denote the 4D energy-momentum tensor by $T_{\mu\nu}$.
The usual scaling symmetry (\ref{full-s})
corresponds to the statement that
$T^{\mu}_{\mu} = 0$.
This is no longer true for us.
The Noether current of the new scaling symmetry
(\ref{scaling-sym}) is
\be
J^{\mu} \equiv \d x^{\nu} T_{\nu}^{\mu}, \qquad
\d x^{\underline{\mu}} = a_{\underline{\mu}} x^{\underline{\mu}},
\ee
where underlined indices are not summed over and
$ a_{\mu}=(a_+, a_-, a_2, a_3) = (0,2,1,1)$.
The corresponding conservation law
$\nabla_{\mu} J^{\mu} = 0$
then implies that
\be
%\sum_{\mu} a_{\mu} T^{\mu}_{\mu} - h T^+_- = 0.
(\nabla_{\mu}\delta x^{\nu})T^{\mu}_{\nu} = 0.
\ee
To derive it,
we have used the energy-momentum conservation law
$\nabla_{\mu}T^{\mu}_{\nu} = 0$.
It would be interesting to derive the trace anomaly for our SYM
theory, which would provide a valuable test to the correspondence we
proposed. 
We expect that some of the techniques for quantum field theory
in curved spacetime can be used here and the problem may
become tractable at least if one treats $h$ and $\chi$ as small perturbations.

Apart from its possible role in holographic duality, by itself the time
dependent super Yang-Mills theory introduced in this paper is already a
very interesting field theory
because of the time-dependent  
gauge and axion couplings
and its untypical scaling symmetry \eq{scaling-sym}.
The new scaling symmetry suggests us to look for a new renormalization group
different from the usual definition associated with a uniform scaling
in all dimensions.
Comparing it with the scaling symmetry of
our supergravity solutions \eq{scaling-sugra},
we see that this new renormalization group should be
one which has a dual interpretation on the supergravity side.
Further work is necessary in order to elucidate these aspects in more
detail. 

The class of solutions which gives a geodesically incomplete spacetime
is of interest for the studies of cosmological singularity. From our
preliminary analysis above of the two-point functions, it is suggested 
that the spacetime structure as seen from the gauge theory is
different from that seen by the classical gravity solution. In fact
from the gauge theory point of view,
the spacetime appears to be non-singular.
It is important to analyze the quantum gauge theory in greater details 
(particularly by including interactions) in order to be
more confident about
this picture and also to learn about how the spacetime singularity is
resolved from a spacetime point of view, and to see 
what kind of interesting 
structures (e.g. quantum symmetries) may appear on the way. 

This paper is only the first step in establishing the connection
between time-dependent backgrounds in string theory and gauge theory.
There are still many important open questions.
For example, as we mentioned at the end of sec. \ref{DBI},
the matching of functional parameters for the duality
may be corrected by terms of higher order in $\alpha'$.
More importantly, we
hope to be able to
understand better the quantum properties of the gauge theory
and use it to learn
about time-dependent processes in
string theory.

%%%%%%%%%%%%%%%%%%%%%%%%%%%%%%%%%%%%%%%%%%%%%%%%%%%

\section*{Acknowledgements}   

We would like to thank Harald Dorn, Clifford Johnson, Veronika Hubeny,
Feng-Li Lin, Don Marolf, Simon Ross, 
Rodolfo Russo, Adam Schwimmer, Shunsuke Teraguchi,
Stefan Theisen, Wen-Yu Wen and Marija Zamaklar 
for discussions. CSC also would like to
thank the string theory group of the National Taiwan University
and the National Center of Theoretical Science, 
of the Humboldt university and of
the Albert Einstein Institute at Postdam for hospitality and
enjoyable discussions. 
The work of CSC was partially supported by EPSRC and PPARC.
The work of PMH is supported in part by
the National Science Council, the National Center for Theoretical
Sciences (NSC 94-2119-M-002-001), Taiwan, R.O.C. and the Center for
Theoretical Physics at National Taiwan University.

%%%%%     %%%%%     %%%%%     %%%%%     %%%%%     %%%%%     %%%%%     %%%%%    
%%%%%     %%%%%     %%%%%     %%%%%     %%%%%     %%%%%     %%%%%     %%%%%

\end{document}